\documentstyle{article}

\begin{document}

\newtheorem{defi}{Definition}
\newcommand{\bdf}{\begin{defi}}
\newcommand{\edf}{\end{defi}}
\newtheorem{prop}{Proposition}
\newcommand{\bpr}{\begin{prop}}
\newcommand{\epr}{\end{prop}}
\newtheorem{example}{Example}
\newcommand{\bex}{\begin{example}}
\newcommand{\eex}{\end{example}}
\newcommand{\pr}{\ni {\em Proof: }}
\newcommand{\rf}[1]{{\rm (\ref{#1})}}
\newtheorem{lemma}{Lemma}
\newcommand{\blm}{\begin{lemma}}
\newcommand{\elm}{\end{lemma}}

%

 \def\ks{Kerr-Schild }
 \def\st{spacetime }
\def\sts{spacetimes }

\def\beq{\begin{eqnarray}}
\def\eeq{\end{eqnarray}}
\def\ben{\begin{enumerate}}
\def\een{\end{enumerate}}
\def\ul{\underline}
\def\ni{\noindent}
\def\nn{\nonumber}
\def\bs{\bigskip}
\def\ms{\medskip}

\def\ab{\alpha \beta}                                        

\def\lh{\hbox{{\boldmath \hbox{$ \ell $}}}}  
\def\lv{\vec \ell }                                                
\def\mh{\hbox{{\boldmath \hbox{$ m $}}}}   
\def\ph{\hbox{{\boldmath \hbox{$ p $}}}}     
\def\qh{\hbox{{\boldmath \hbox{$ q $}}}}      
\def\uh{\hbox{{\boldmath \hbox{$ u $}}}}      
\def\nh{\hbox{{\boldmath \hbox{$ n $}}}}      
\def\ll{\lh \otimes \lh }                                   
\def\mm{\mh \otimes \mh }                           
\def\pp{\ph \otimes \ph }                               
\def\qq{\qh \otimes \qh }                               
\def\uu{\uh \otimes \uh }                               
\def\nn{\nh \otimes \nh}                                
\def\slm{\lh \otimes \mh + \mh \otimes \lh} 
\def\spq{\ph \otimes \qh + \qh \otimes \ph}  
\def\sun{\uh \otimes \nh + \nh \otimes \uh}  

\def\lv{\vec l }                                     
\def\mv{\vec m }                                     
\def\xiv{\vec \xi}                                   
\def\lie{{\cal L}({\scriptstyle \xiv}) }             
\def\ll{\lh \otimes \lh }                           

\def\of{{\bf \hbox{$ \Theta $}}}          
\def\ofo{ {\of}^{\Omega}}      
\def\ofl{ {\of}^{\Lambda}}  
\def\ofp{ {\of}^{\Pi}}              
\def\tol{\of\sp{\Omega} \otimes \of\sp{\Lambda} } 
\def\cb{ \{ {\of}^{\Omega} \} }           
\def\bg{{g}}  	                    
%
%
%
%
%
\title{A physical application of Kerr-Schild groups}
\author{Sergi R. Hildebrandt\cr
{\small Instituto de Ciencias del Espacio (CSIC) \&}\cr
{\small Institut d'estudis Espacials de Catalunya (IEEC/CSIC)}\cr
{\small Edifici Nexus, Gran Capit\`a 2-4, 08034 Barcelona, Spain}\cr
{\small hildebrandt@ieec.fcr.es, http://www.ieec.fcr.es/cosmo-www/}}
\maketitle
\date{}
\begin{abstract}
\baselineskip.34mm
The present work deals with the search of useful physical 
applications of some generalized groups of metric transformations.
We put forward different proposals and focus our attention on 
the implementation 
of one of them. Particularly, the results show how one can control 
very efficiently the kind of \sts related by a Generalized 
Kerr-Schild (GKS) Ansatz through Kerr-Schild groups. 
Finally a preliminar study regarding other generalized 
groups of metric transformations is undertaken which 
is aimed at giving some hints in
new Ans\"atze to finding useful solutions to Einstein's equations.
 \end{abstract}

\section{Introduction}

Kerr-Schild groups (or motions) are new kind of groups of metric transformations \footnote{The transformations are only required to be 
well behaved locally and  the words ``group'' or ``transformation'' 
will stand for ``local group'' or ``local transformation''.}.
In \cite{ksvf}, their general structure as (local) Lie groups is worked out and several explict solutions are solved. In another work, \cite{ksgmm},  a further study has been carried out  to deal with the issue of their general resolution and existence. The results show that \ks groups have a much richer structure than isometries or conformal symmetries. Some of the (new) features worth to be remarked are that they introduce a null vector field as the basic ingredient for the definition of a metric transformation, under some circumstances they contain infinite dimensional Lie algebras, or they allow for restricting isometries into subgroups according to their relation with respect to the null vector field. In this paper we shall 
start from the result that \ks motions may be characterized 
by the system of differential equations (\ks equations 
hereafter)
\beq
\label{eq-ks}
 \lie g   =  2h \ll, \qquad \lie \lh   =  b \, \lh ,
\eeq
where $ g $ is the fundamental tensor of a (Lorentzian) 
manifold, and $ \lh $ is a real-valued one-form field. These objects 
are considered as data of the above system.\footnote{Only the direction of $ \lh $ is relevant. 
Therefore, it is worth to refer to
the congruence of null curves with tangent vector 
$ \lh $, say $ {\cal C}_{\ell} $.} 
On the other hand, $ h $ and $ b $ are unknown $ C^{\infty} $ 
functions and $ \xiv $ is the infinitesimal generator of a \ks 
motion, also an unknown of the problem. Particularly, $ \xiv $ 
is named a ``\ks vector filed'' (KSVF) in the same way that we 
have Killing or conformal vector fields.

In addition, other works in \cite{ksgmm} we have tackled other new possibilities coming from a geometrical definition of metric transformations.
That work has enough detailed and general results. However their physical application is not developed in accordance.

In this work we give some ideas that may prove to be useful towards
such goal. Some of them are  based upon physical applications of the 
two well-known metric symmetries, i.e. isometries and conformal symmetries. 
The main point here is that \ks groups, or the new studied candidates, may yield 
relevant results beyond the ones of the two mentioned symmetries 
in some areas. Our aim is mainly to bring the attention to them.

In Sect.~\ref{sec-som}, some applications are put forward. 
Sect.~\ref{sec-app} defines one of the presented ideas, focussing the attention 
on the \ks case, because it is, as of now, the most 
studied situation in the literature. 
In Sect.~\ref{sec-geo} we apply it to geodesic \ks groups 
(\ks groups in which $ \lh $ is geodesic).
In Sect.~\ref{sec-fir}, we begin with new candidates to metric 
transformations (or generalized metric symmetries) and give the 
results for the signature of the obtained \sts (following the idea of 
Sect.\ref{sec-app}) in order to discern which may be the most 
interesting cases to be further studied elsewhere. We end this 
paper with the Conclusions.
\section{Some physical applications of generalized \\metric symmetries}
\label{sec-som}
As mentioned elsewhere, their use in physics is just starting. Therefore, we can only advance some ideas about their application in
direct physical problems. These will be surely complemented in the future, 
but we shall show that there are already some hints.
First, there are some works which implement usual 
metric symmetries (isometries and conformal symmetries) to 
Einstein's equations, see for instance \cite{zafiris2}-\cite{archel}.
This implementation has yielded interesting results because the authors 
collected  different theorems and propositions on the energy-matter content that were previously a part, as well as new results. But the most remarkable feature would be the direct way they were all treated together thanks to the use of isometries 
or conformal transformations. 
We believe that a similar work might prove to be as useful when based upon other sort of metric symmetries and will be dealt with elsewhere.

Secondly, even in the field of the problem of rigidity and elasticity in General Relativity, there appear some problems that can be written in terms of metric symmetries, or affine-like ones, see e.g. \cite{bona}-\cite{ere96}.
For instance, the Beltrami-Michel equations for any stress tensor, \cite{bl}, can be obtained from the integrability conditions of a generalized metric symmetry. Moreover, each of the proposals that aim at generalizing the classical group 
of rigid motions in General Relativity could benefit from their analysis inside the frame of generalized symmetries. An example regards to the appearance of 
infinite dimensional algebras in our scheme which might be linked with 
the action of a certain group, or subgroup, of the generalized rigid motions. 
This will be the matter of subsequent work.

Finally, there is another possibility, closer to the main line of thought taken 
hitherto in the literature. The idea of using metric symmetries in order to group 
\sts is very well-stablished, specially for the case of isometries, see e.g. \cite{kramer,petrov}. In fact it is nowadays customary to use expressions such as $ 
G_3 $, $ G_6 $, \ldots Therefore, with the addition of new metric symmetries, we could begin to use similar classifications. 
Yet we shall focus on a refined idea:

Instead of just classifying \sts taking into account only inner metric symmetries, 
that would constitute a trivial extension of the usual isometric classification, we demand moreover that there exists some ``external'' relation among them. 
Of course, there are many alternatives for these ``{\it adhoc}'' requirement, 
but we have decided that the very metric symmetry be the connection amongst such 
spacetimes. This is accomplished by demanding that the \sts that share the same algebra 
of a given metric symmetry must be linked to each other by finite relations of the
 same form as the considered local metric symmetry. The meaning of being of the same form will be defined in a moment.

This idea will be mainly applied to the \ks case in this work because 
the (Generalized) \ks (GKS) Ansatz has proven to be very useful in General Relativity,
see e.g. \cite{kramer,ks,gks}. The fact is that, although the same idea can be further applied  without difficulty to other new metric symmetries, one should first discern which are the most interesting 
finite metric relations. One often asks for new Ans\"atze that would as well yield 
useful solutions to Einstein's equations. However, their analysis 
turns out to be rather cumbersome for almost any alternative 
to conformal or GKS ones, see e.g. \cite{plebans}. 
Now, we have the bonus that the structure of the final metrics ultimately rests 
on a symmetry problem, which, thanks to Lie's theory on continuous groups, is entirely controlled in its linear and first order contributions, also gives the hope that other metric transformations, e.g. the ones in~\cite{ksgmm}, will prove to be useful.
 We had not still carried out an exhaustive study for these much bigger set of new metric symmetries. Here  we present only a basic step towards such goal in Sect.~\ref{sec-fir}, which are the signatures of the resulting \sts and if they may change.

\section{Application to \ks groups}
\label{sec-app}
%
%
\subsection{A preliminary remark}
Among the results obtained in \cite{ksvf} we would like to stress the following one
\bpr
\label{prop-ks}
If a \st is of a GKS form, i.e. $ {\tilde g} = \bg + 2 H \ll $, the solution 
for its KSVFs
is the same as for $ (\bg, \lh ) $, i.e. Eqs.~\rf{eq-ks}, where now 
$ {\tilde b} = b $, $ {\tilde h} = h + \lie H + 2 b H $ .
\epr
It is important to notice that the set $ \{ \xiv \} $ is the same, but not necessarily its action on $ \tilde g $, since, in general, $ {\tilde h} \neq h $. This proposition 
will be useful in the following section and the examples later on.
\subsection{Integration {\em \`a la Kerr-Schild} of a \ks group}
\bdf[Integration {\em \`a la Kerr-Schild}]
\label{def-int}
We will say that we have integrated {\em \`a la Kerr{-}Schild}\/ 
some particular \ks group whenever
\beq
\label{eq-int}
 \exists H \ \big| \ \lie H + 2 H b + h \in {\cal F}( h ),\qquad \forall \xiv ,
\eeq
where $ H $ is a $ C^{\infty} ${-}function of the manifold and $ {\cal F}( h ) $ is the set of functions that have the same functionality in the variables of the manifold as $ h $ has, and $ \xiv $ is any KSVF of the considered \ks
group.
\edf
Consequently, from Proposition~\ref{prop-ks}, and Eq.~\rf{eq-int}, $ {\tilde g} \equiv \bg + 2 H \ll $ (where $ \lh, b, \xiv, h $ are the same objects of the considered \ks problem) will have not only the {\em same} \ks group as $ g $, $ \lh $, but also its action on $ {\tilde g} $, $ g $. The same could 
 be straightforwardly extended to other kind of general metric symmetries. 
 Let us recall that the notion seems by itself interesting: in order 
to study a particular finite 
metric relation, which is non-linear and, often, difficult to handle, use 
moreover the information given by its ``corresponding'' infinitesimal 
transformation, which is linear and of first order. Note that this analysis 
would not make any sense in the case 
of isometries, since the subset of metrics which are related 
by an isometric relation is simply the same metric itself. 

Since $ \lh $ is a null one-form field, only its direction is relevant 
in the previous definition, that is
\bpr
The {\em \`a la Kerr-Schild} integration does not dependend on the parametrization of $ \lv $.
\epr
\pr If the parametrization of the curves defining the vector field $ \lv $ 
is changed, $ \lv $ changes by a multiplicative factor, i.e. 
$ \lv \to \lv' = A \lv $, with $ A \neq 0 $ and henceforth 
$ \lv \to \lh' = A \lh $. Writing the problem of \ks groups for 
two different parametrizations, eqs.~\rf{eq-ks}, it turns out that
\beq
h' = h/A^2, \qquad b' = b + \lie \ln |A|,
\eeq
where $ h $ and $ b $ are the solutions for a certain parametrization and $ A $ ($\neq 0$) measures the change of parametrization between both cases. Then, eqs.~\rf{eq-int} for $ \lh' $ are $ {\tilde h}' = h' + \lie H' + 2 b' H' \in {\cal F}(h') $. 
The solution of these equations is clearly $ H' = H / A^2 $, where $ H $ is the solution to $ A = 1 $. Finally, writing $ g' = g + 2 H' \ell' \otimes \ell' $, one gets $ g' = g + 2H \ll $. Thus, the only characteristic property of $ \lh $ is its direction, as remarked elsewhere.

We now proceed to solve the most representative examples worked out in \cite{ksvf} as well as a new one, which is the general solution of \ks groups
 for Kerr-Newman \sts taking $ \lh $ to be one of its principal null directions, Example~\ref{ex-ker}. The reason of the subdivision in the presented examples is to be found in the results on the study of the general existence
of \ks groups, \cite{ksgmm}.  The non-geodesic case is left aside since the obtained solutions in \cite{ksgmm} do not 
 have apparently an immediate physical use.
\section{Geodesic $ \lh $ with $ \Delta \neq 0 $}
\label{sec-geo}
To begin with we define the {\it scalar} $ \Delta $ as:
\beq
 \Delta := - 2 D \theta + 4 \theta^2 - 3 {\bar R} + 2 l^{\mu} \nabla_{\sigma} 
\nabla^{\sigma} l_{\mu} + DM - 2M \theta \neq 0 ,
\eeq
where 
\beq
\theta = {1 \over 2} \nabla_{\rho} l^{\rho}, \quad 
D := l^{\rho} \nabla_{\rho}, \quad {\bar R} = R_{\rho \mu} 
l^{\rho} l^{\mu}, 
\eeq
and $ M $ is defined by $ \lh $ geodesic $ \Longleftrightarrow  {\vec a} \equiv D \lv = M \lv $. 

We include it here because this scalar appears as a fundamental quantity in the study of the general existence of \ks groups. However, in the present work, it is only necessary to know that it classifies the possible \ks groups generated by geodesic $ \lh $, according to whether $ \Delta $ vanishes or not..
\bex[Radial and spherically symmetric $ \!\lh $ in flat \st$\!\!$]
\label{ex-rad} 
\hspace{-.33cm} The integration {\em \`a la} \ks for any geodesic, radial, spherically symmetric null vector field of flat \st yields the set of \sts linked by $ \bg  = {\bf \eta} + 2 H \ll $.
\eex
This first example of this section comes from flat spacetime, but one would have attained the same result starting from any of the resulting $ \bg $.

\pr The \ks equations read
\beq
\label{eq-ksp}
\lie \eta = 2 h \ll , \qquad \lie \lh = b \lh .
\eeq
The solution to this problem taking $ \lh = { 1 \over  \sqrt{2}}d(t \pm r) $ , where we use standard spherical coordinates
$$ ds^2 = -dt^2 + dr^2 + r^2(d\theta^2 + {\sin^2}\theta \, d\varphi^2), $$
is shown to be $ h = 0 $, $ b = 0 $ and $ \{ \xiv \} = SO(3) \otimes T_{t} $, 
where $ T_{t} $ stands for translations along the $ t $ axis.

The equations that we need to solve are then
\beq \lie H = 0 
\eeq
whose solution is $ H = H(r) $. Thus, the collection of metrics related by a GKS transformation that have $ SO(3) \otimes T_{t} $ 
as a solution to their own spherical \ks problem is
\beq {\tilde g} = \eta + 2 H(r) \ll \, , 
\eeq
with $ \lh ={1 / \sqrt{2}}( dt \pm dr )$.

The set of metrics read in these coordinates (we focus on the case $ \lh = (1/\sqrt{2})(dt + dr) $, though the same spacetiems are obtained for $ \lh = (1/\sqrt{2})(dt - dr) $)
\beq
ds^2 = -(1-H)dt^2 + 2H dt\,dr + (1+H)dr^2 + r^2d\Omega^2, \quad H=H(r) .
\eeq
A general cobasis, which avoids coordinate problems at the possible horizons, is
\beq
\begin{array}{ll}
 \of\sp{0}  =  ( 1- {H \over 2} ) dt - {H \over 2} dr ,\quad &
 \of\sp{1}  =  ( 1+{H \over 2} ) dr + {H \over 2} dt, \cr
 \of\sp{2}  =  r \ d\theta, \quad & \of\sp{3}  =  r \sin \theta \ d\varphi . \cr
\end{array}
\eeq

One can recover for $ H < 1 $ the explicit static expression of the metric by making the well-known change of the $ t $ coordinate given by $ dt = dt' + [H /(1-H)]\,dr $,
where $ t' $ is a new coordinate, whereas $ r $, $ \theta $, and $ \phi $ remain unchanged. In this set of coordinates one gets
\beq
\label{eq-hl1}
ds^2 = - (1- H )\, (dt')^2 + (1 - H)^{-1}\, dr^2 + r^2\, d\Omega^2, \quad H=H(r) .
\eeq
We change in the following the name of $ t' $ to $ t $, the usual Schwarzchild time coordinate. The natural cobasis to study all these metrics in the region $ H < 1$ is then $ \of\sp{0}  =  \sqrt{1 - H}\,dt $, $ \of\sp{1} = (1/\sqrt{1 - H}) dr $, 
$ \of\sp{2} =r \, d\theta $, $ \of\sp{3} = r \sin \theta \, d\varphi $.

The Riemannian tensor has the following non-zero components (and, in addition,
 the ones obtained by index symmetries). The expressions are equally valid for both cobasis.\footnote{This result can be extended to any other changed cobasis in any of the two two-planes expanded by $ dt-dr $ and $ d\theta-d\varphi $.} 
\beq
\begin{array}{l}
 R_{0101} = - {{H''}/ 2} , \quad R_{0202} = R_{0303} = -{{H'} / 2r} , \cr
\ R_{1212} = R_{1313} = {{H'} / 2r} \quad R_{2323} = { H /  r\sp{2}}.
\end{array}
\eeq

The Ricci tensor has the following non-zero components
\beq
R_{00} = -R_{11} = -{1 \over 2}\left( H'' + {2 H'\over r} \right), \quad R_{22} = R_{33} = {1 \over r}\left( H' + {H \over r} \right) .
\eeq
And the scalar curvature is given by
\beq
R = H'' + {4 H' \over r} + {2 H \over r\sp{2}} . 
\eeq

Finally, their Einstein tensor has the following expression
\beq
G_{00} = -G_{11} = { 1 \over r} \left( H' + { H \over r} \right) , \quad G_{22} = G_{33} = - {1 \over 2}\left( H'' + {2 H'\over r} \right) .
\eeq
Among the $ (V_4, {\tilde g}) $ one finds Schwarzschild vacuum and interior 
solution, Reiss\-ner-Nordstr{\"o}m and de Sitter solutions, and many others. 
Particularly, these are obtained with ($ G = c = 1 $ throughout this work) 
$ H = 2m/r $ (the only non-flat vacuum solution within the above set), 
$ H = 2m/r - Q^2/r^2 $, $ H =(\Lambda/3) \, r^2 $, respectively, and 
where $ m $ is the mass of the object, $ Q $ its charge and $ \Lambda $ is the cosmological constant. The only case going back to flat \st is the trivial one 
$ H = 0 $. In order to show the usefulness of the preceding results, we will say that 
they have been applied very recently to study (quantum) regular interiors for black 
holes, see e.g. \cite{magli}-\cite{rqibh}.

Besides these conclusions, it is worth emphasizing that the appearance of the de Sitter's solution has been in the form of its {\em causal connected part}. 
Thus this result may add some hints about the role played by 
$ \lh $ (this result must be related to the fact that $ \lh $ is a null one-form) in the \ks transformation and its physical meaning, see also \cite{coll}. Results which, 
on the other hand, may as well be useful for the new situations depicted in 
Sect.~\ref{sec-fir}, in order to relate the desired physical properties of the final solutions with the type of transformation one has to use for.
\bex[Kerr-Newman \sts]
\label{ex-ker} The integration {\em \`a la} \ks for any of the principal null directions of Kerr-Newman \sts is given by $ \bg  = {\bf \eta} + 2 H(r,z) \ll $, if $ \lh $ is rotational and by the result of Example~\ref{ex-rad} for the case of an irrotational $ \lh $.
\eex

The first thing to do is to solve the \ks groups in Kerr-Newman \sts for their principal null directions. The result is
\blm
\label{lem-kn}
The \ks groups for Kerr-Newman spaces associated with the principal null directions are given by $ T_{t} \otimes T_{\phi} $ for rotational $ \lh $, and by $ T_{t} \otimes SO(3) $ for the irrotational case.
\elm
Here $ t $ and $ \phi $ are Kerr coordinates, not Boyer-Lindquist coordinates. 
This result is remarkable because it extends \ks groups to the most useful models of charged, spinning objects.

\pr The path of solving eqs.~\rf{eq-int} directly is unnecessarily long. 
Instead we can make use of Proposition~\ref{prop-ks}.

In our case, it is well-known that Kerr-Newman \sts can be written in a \ks form, 
i.e. $ \bg = \eta + 2 H \ll $, where $ \lh $ is any of its principal null directions, and 
$ H = H(r,z) $, see e.g. \cite{kramer,mtw}, yet its expression is irrelevant to our aims. The consequence ---using Proposition~\ref{prop-ks}--- is that the 
\ks problem for Kerr-Newman 
\sts chosing $ \lh $ to be one of its principal null directions can be reduced to the resolution of Eqs.~\rf{eq-ksp}, where $ \lh $ has the expression
\beq
\label{eq-l}
\lh = {1 \over \sqrt{2}} \biggl( dt + {z \over r} \, dz + {rx + ay \over r^2 + a^2} \, dx + {ry - ax \over r^2 + a^2} \, dy \Biggl),
\eeq
where $ a $ is now simply a constant, (its meaning as the spin of a particle belongs to Kerr-Newman \sts) and $ \{ t, x, y, z \} $ are now cartesian coordinates of
the flat spacetime. It is easy to show that $ \lh $ is geodesic and shear-free. 
If $ a $ is zero, we recover the radial null vector of Example~\ref{ex-rad}. 
If $ a \ne 0 $, $ \lh $ is rotational.  The function $ r $ is defined by 
$ r^2(x^2+y^2+z^2) +a^2z^2 = r^2(r^2+a^2) $ as in Kerr-Newman spacetimes
and the $ 1/\sqrt{2} $ factor is introduced in order to link the 
irrotational situation with the radial $ \lh $ of Example~\ref{ex-rad}.

Again, a  direct resolution of Eqs.~\rf{eq-ksp} is rather long, though easier 
than a direct resolution within Kerr-Newman spacetimes. Nevertheless, we shall 
make use of a ---non-trivial--- result about the general resolution of \ks groups.
When $ \lh $ is geodesic, it can be shown that $ \Delta $ determines the type of solution to the problem of \ks groups. Particularly, if $ \Delta \neq 0 $, 
KSVFs are the solutions of the system
\beq
\label{eq-gam}
\lie \gamma = 0, \quad \lie \lh = b \lh , 
\eeq
where $ \gamma $ is a rang-two symmetric tensor, whose exact expression is unnecessary now, see \cite{ksgmm} for details. It suffices to know that, in our case, we have the bonus that $ \gamma = \eta $. Thus our problem reduces 
to a problem of isometries {\em restricted} by the second set of 
equations in~\rf{eq-gam}.

In the same reference, one can find the conditions under which $ \Delta $ 
may vanish in, e.g., flat spacetime. A condition is that $ \lh $ cannot 
be rotational. Therefore for $ a \neq 0 $, $ \Delta \neq 0 $. 
For the irrotational case, $ a = 0 $, we can directly compute 
$ \Delta $ since the expressions are rather simple. The result is $ \Delta = 1/r^2 \neq 0 $.
Thus $ \Delta \neq 0 $ for the principal null directions of Kerr-Newman metrics, regardless the value of $ a $.

The final step is to solve Eqs.~\rf{eq-gam}. The solutions must be a linear
 combination of the infinitesimal generators of the Poincar\'e group that satisfy
the second group of equations of~\rf{eq-gam}. The calculations are given 
in \cite{ksgmm}. The end result is that $ h = b = 0 $, and $ \{ \xiv \} = \{ \partial_t, x \partial_{y} - y \partial_{x} \} $ for the case $ a \neq 0 $, and $ \{ \partial_t, x\partial_y - y \partial_x, x \partial_z - z \partial_x, y \partial_z - z \partial_y \} $. This is the solution for flat spacetime. The infinitesimal generators are the same for  Kerr-Newman spaces as we know from Proposition~\ref{prop-ks}.
 We can also use it in order to find their action on them. The result is, in any case, $ \tilde h = \tilde b  = 0 $. Therefore, the KSVFs are the Killing vectors of 
Kerr-Newman spaces. This finishes the proof of Lemma~\ref{lem-kn}.

The final step is to solve Eqs.~\rf{eq-int}. The solution is: for the rotational case,
$ H = H(r,z) $ and for the irrotational case, $ H = H (r) $.

We note that a restriction of the \sts corresponding to the rotational case has been
recently used in \cite{burinskii} in order to deal with the problem of a (quantum) 
source origin for the (charged) spinning particle within supergravity and 
string fields for the source. The new spacetimes are chosen from a direct generalization of Kerr-Newman spacetimes, letting the mass function 
to have a free function of $ r $. We remark taht the whole family, above presented, 
has not been used yet. We reckon that taking it into account, and using the local irrotational set of observers in order to set the physical conditions near the core might yield new results ---recall the results in the radial (irrotational) case before---, specially for the case where vacuum polarization is
the dominant effect.

Finally, let us recall that the main idea behind the GKS Ansatz is to choose a 
particular seed metric, say $ g_0 $, and an $ \lh $ staisfying some physical
requirements. This procedure is stillpresent in \ks groups because 
$ g $ and $ \lh $ (its direction) are the data of the problem. One is thus 
encouraged to perform a further study for other geodesic $ \lh $ with 
$ \Delta \ne 0 $. For instance, we point out that Vaydia 
and Kerr-Vaydia metrics, \cite{kv}, are spacetimes which could benefit from \ks groups for the obtention of some hints towards interior solutions. Moreover, another option
could start from the solution of \ks symmetries for static spherically symmetric \sts given in \cite{ksvf}. Besides the ``$ \rho + p = 0 $''  family, the rest of interior stellar models would be equally subgrouped according to their shared \ks symmetries which should lead to similar results to those of \cite{magli,rqibh}. Finally, the connection 
with cosmological issues can be started from \cite{ss}.
\subsection{Geodesic $ \lh $ with $ \Delta = 0 $}
This is the last section on KSVFs. The null vector fields which are
 geodesic and satisfy $ \Delta = 0 $ are still a matter of further research. 
For the time being, it suffices to know that in spacetimes where 
$ \lv $ is a principal null direction and $ \bar R = 0 $ the family of such 
vectors includes the ``cylindrical'' and ``cartesian'' (or ``parallel'') cases, see below. 
Both cases constitute paradigms of \ks groups. For the sake of brevity, we shall 
consider flat \st in the examples, although an extension 
of the results to several spacetimes of physical 
interest, such as those representing {\em pp}-waves, is readily 
accomplished (see later). Using cylindrical coordinates, the line-element 
of flat spacetime takes the form
\beq
ds^2 = -\,dt^2 +\, d\rho^2 + \rho^2\, d\varphi^2 +\, dz^2 .
\eeq
The ``cylindrical'' $ \lh $ corresponds to $ \lh_{(\pm)} = (1/\sqrt{2})
(\pm dt + d\rho) $. In the second case, using cartesian coordinates,
\beq
ds^2 = - \, dt^2 + \,dx^2+\,d y^2 + \, dz^2,
\eeq
whence one defines $ \lh_{(\pm)} = (1/\sqrt{2})(\pm dt + dx) $ as a ``cartesian'' 
(or ``parallel'') $ \lh $ (of course, $ x $ may be interchanged 
by either $ y $ or $ z $).

Both $ \lh $ yield rather particular solutions to the problem of \ks
groups. In the first case, we find (see \cite{ksvf}) a local group which must be reduced when considering global topological properties of flat spacetime, 
whereas in the cartesian case, the Lie algebra turns out to be infinite dimensional, i.e. containing some {\em functional} freedoms. 

Due to these special properties, it is worth studying them here. The results are
\bex
\label{ex-cyl}
The {\em \`a la} \ks integration of the cylindrical $ \lh $ in flat spacetime
yields the \st defined by the metric tensor 
$ g = \eta + 2 H(\rho) \ll $, where $ \lh = (1/\sqrt{2}) (\pm dt + d\rho) $, and $ \{ t, \rho \} $ are usual cylindrical coordinates of cylindrically symmetric spacetimes.
\eex
The calculation of this result follows similar steps as those of the preceding 
examples. The expressions for the KSVFs can be read from \cite{ksvf}. 
We note that the result above is valid either for the local 
group and for the global one.

The geometrical properties of this case can be worked with the aid of two 
different orthonormal cobasis as before. An analogous change in the time 
coordinate $ t $ as that of the previous section allows us to write them in an explicit static form if $ H < 1$. The expressions are analogous to~\rf{eq-hl1}.

The Riemann tensor components are \footnote{The results are again the same for several cobasis, including the ``regular'' and the ``static´´ ones.} ($ H= H (\rho) $, and $ A' \equiv dA/d\rho $)
\beq
R_{0101} = - {H'' / 2} , \quad R_{0202} = R_{0303} = -{H' / 2\rho} , \quad R_{1212} = {H' / 2 \rho},
\eeq
and the rest are obtained by index permutation or are zero. 
The Ricci tensor for these metric-spaces has the following non-zero components
\beq
 R_{00} = -R_{11} = -(1 / 2)[ H'' + ( H'/\rho)], \quad R_{22} = {H' / \rho} .
\eeq
Whence the scalar curvature is given by $  R = H'' + (2 H' / \rho) $.

Finally, their Einstein tensor has the following non-zero components.
\beq
G_{00} = -G_{11} = { H' / 2 \rho} , \quad G_{22} = - {H'' / 2}, \quad G_{33} = G_{22} + 2 G_{11} .
\eeq
The addition of a cosmological constant does not change the relation between 
$ T_{00} $ and $ T_{11} $, though the relation among the spatial pressures allows for changes in the signs of $ T_{22} $ and $ T_{33} $. The situation is now different with respect to Example~\ref{ex-rad}, where we had well-known solutions for the classical vacuum case, or the case of electromagnetic fields. We have now an axis of symmetry. The only vacuum solution is now flat spacetime,
 which is obtained with $ H = const. $. The solution for $ R  = 0 $ is $ H = a + b/\rho $, where $ a $, $ b $ are constants ($ a $ is a gauge freedom), and corresponds to the electromagnetic field of a linear charged distribution, located along the axis $ \rho =  0 $, and is an analogue of the Reissner-Nordtr\"om solution for the symmetry we are considering.

We shall not analyze here their geometric properties (singularities, possible extensions, etc). It seems that no well-stablished physical system 
can be attached to such relations and values of pressure and density.

\bex
\label{ex-car}
The {\em \`a la} \ks integration of the cartesian (plane) $ \lh $ in flat 
Spacetime can not be accomplished.
\eex
This result comes from the fact that the Lie algebra is now infinite dimensional as
we will now show. For $ ds^2 = - 2 \, du \, dv + dx^2 + dy^2 $, and 
$ \lh = du $, where $ u := { 1 /  \sqrt{2}}(t + z) $, 
$ v := { 1 / \sqrt{2}}(t - z) $, being $ t $, $ x $, $ y $, $ z $ standard cartesian coordinates of flat spacetime, the system of Eqs.~\rf{eq-ksp} has the following solution:
\beq
\begin{array}{l}
 h =  - {\ddot A} \, x - {\ddot B} \, y - {\dot C} + {\ddot D}  \, v , \quad  b  = {\dot D } \cr
\xiv =  D \, \partial_u + \bigl[  -{\dot F}\, v + {\dot A}\, x + {\dot B}\,  y + C \bigr] \partial_v  +  \bigl( -d \, y + A \bigr) \partial_x  + \bigl( d \, x + B \bigr) \partial_y ,\cr
\end{array}
\eeq
where the functions $ A $, $ B $, $ C $, $ D $ are arbitrary $ C^{\infty} $ functions of the variable $ u $, $ {\dot A} \equiv dA/du $, etc., and $ d $ is a real constant. It is the presence of these free functions that tells us that the associated Lie algebras are infinite dimensional.

Consequently, the integrating system is now
\beq
\label{eq-hti}
 \lie H + 2H {\dot D} = {\tilde h} - h,
\eeq
where $ {\tilde h} = -({\tilde A})\,\ddot{}\, x - \ldots \,$, changing all functions and constants by their respective  expressions with tilde.

Notice that the solution has to be valid for any combination of the unknown 
functions and constants of the \ks solution. The end result is that
only the trivial solution $ H = 0 $ satisfies Eqs.~\rf{eq-hti} and therefore
Definition~\ref{def-int}.

Notice that from Proposition~\ref{prop-ks} the results above can be extended to include any \st with $ g = \eta + F(u,v,x,y) \, du\otimes du $. Particularly, 
if $ \partial F / \partial_v = 0 $ we have the pp-waves metrics, see e.g. \cite{kramer,ek}.

 This result does not mean that for any subgroups the 
{\em \`a la } \ks integration exists, e.g. for any unidimensional 
subgroup. However, it is enough, as of now, to 
remark the impossibility of fulfilling Definition~\ref{def-int} for some 
general (intrinsic) \ks algebras. On the other hand, it yields
not only a limit for the grouping that can be attained with Definition~\ref{def-int} itseld ---forbidding too free solutions---, but it might also help understanding some solutions to Einstein's equations from a different viewpoint. 
For instance, pp-waves and flat spacetime do {\em not} share their KSVFs, 
whereas Kerr-Newman and flat spacetimes {\em do}. A similar result 
was obtained twenty-five years ago by L. Defrise-Carter \cite{def-car} regarding the 
``isometrization'' of conformal symmetries. All 
this adds a new argument to the conclusion that although Kerr-Newman and 
pp-waves are \ks metrics, their relation with flat spacetime is 
completely different.
\section{First steps towards finding other metric symmetries}
\label{sec-fir}
We will now introduce other metric symmetries candidates.
In \cite{ksgmm} some new candidates of metric symmetries were proposed and
some examples were presented. The examples were mainly intented to prove 
the existence of Lie algebras for each candidate. The aim here is to give additional
arguments in favor of their study. We thus think that a knowledge of some of their 
physical consequences could
help deciding which are worth to be studied in more detail. A necessary point 
to be considered is the signature change that may happen when using such new metric 
relations. In the \ks situation this issue was absent because a contribution 
of the type $ \ll $ with $ \lh \cdot \lh = 0 $ does not change its signature. For the rest of the situations ---excluding isometries--- the signature of the final metric may change. Let us finally recall the point stressed 
in the introduction that a non-Lorentzian signature may be of interest if, e. g., 
we aim at dealing with problems where space--like metrics are the 
relevant object, or where signature change plays a definite role.

Of course, for the sake of brevity, we refer the reader to \cite{ksgmm} for details on these metric symmetries.
\subsection{The $ \lh ${-}$ \mh $ candidates}
\label{sec-lmc}
This set is built upon the metric symmetries that a pair of null one-form fields, say $ \lh $ and $ \mh $ with $ \lh \wedge \mh \neq 0 $, can create and is related with the light cone structure of spacetime. Obviously, a subcase would be that of \ks groups
developed before, where only one null vector field is used. 

The associated finite relations read
\beq
{\tilde g} = g + H\, \ll + G \, (\slm) + F\, \mm,
\eeq
where $ F $, $ G $, $ H $ are $ C^{\infty} $ functions on the manifold. 
The study of the signature of $ {\tilde g} $ yields, see also \cite{tm} \footnote{We shall choose a Lorentzian signature for $ g $, particularly, $ g = diag(-1,1,1,1) $. Of course, similar results are valid if one starts with the $ g= diag(1,-1,-1,-1) $.}
$$ \begin{array}{||l|ll|}
\hline
& 4 HF > (1+G)^2, & E, \cr
\hbox{sign ($ H $) = sign ( $ F $ ) = 1}
& 4 HF = (1+G)^2, & D ,\cr
& 4 HF < (1+G)^2, & L.\cr
\hline
 & 4 HF > (1+G)^2, & A, \cr
\hbox{sign ($ H $) = sign ($ F $) = -1} & 4 H F = (1+G)^2, & D,\cr
& 4 HF < (1+G)^2, & L.\cr
\hline
\hbox{sign ($ H $) $ \neq $ sign ($ F $)}  & L.  & \cr
\hline
\hbox{$ H = 0 $ or $ F = 0 $} & L.  \hbox{ For $ G = -1$,} & D.\cr
\hline
\hbox{$ H = 0 $ and $ F = 0 $} & L.  \hbox{ For $ G = -1$,} & \hbox{double }D.\cr
\hline
\hline
\end{array} $$

Here (E)uclidean, (D)egenerated, (L)orentzian, (A)rtinian and 
double (D)e\-generated signatures mean $(1,1,\allowbreak 1,1)$, 
$(0,1,1,1)$, $(-1,1,1,1)$, $ (-1,-1,1,1) $, \allowbreak$ (0,0,1,1) $ 
respectively.
\subsection{The $ \ph ${-}$ \qh $ candidates}
\label{sec-pqc}
This set is a similar one with respect the previous Section. The difference 
lies now in that the metric symmetries are built using a pair of space-like 
one-form fields, say $ \ph $ and $ \qh $ with $ \ph \wedge \qh \neq 0 $. 
Their corresponding finite transformations are
\beq
 {\tilde g} = g + H \, \pp + G \, (\spq) + F \, \qq . 
\eeq
The study of the signature of $ \tilde g $ yields
$$ \begin{array}{||l|ll|}
\hline
& G^2 > (1+H)(1+F), & A.\cr
\hbox{sign($H+1$) = sign($F+1$) }= \pm 1 & G^2 = (1+H)(1+F), & D.\cr
& G^2 < (1+H)(1+F), & L, \cr
\hline
\hbox{sign ($ H + 1 $) $ \neq $ sign ($ F + 1 $)}  &   A. & \cr
\hline
\hbox{$ H = -1 $ or $ F = -1 $} & L. \hbox{ For $ G = 0 $,} & D. \cr
\hline
\hbox{$H=-1$, $F=-1$, $ G = 0$}  & \hbox{ double D.} &\cr
\hline
\hline
\end{array} $$

In this case Artinian, Degenerated and double Degenerated signatures refer to $ (-1,1,1,-1)  $, $ (-1,1,1,0) $ and $ (-1,1,0,0) $, respectively.
\subsection{The $ \uh ${-}$ \ph $ candidates}
\label{sec-upc}
This last set is built upon the metric symmetries that a time-like one-form field and a space--like one, say $ \uh $ and $ \nh $ with $ \uh \wedge \nh \neq 0 $ can create. Of course, combinations of this group correspond to different cases of Section~\ref{sec-lmc}. Nevertheless, there some problems where it is easier to work directly with a time-like vector field directly. For insance, in a ``time-like'' version of the GKS
Ansatz, that is, if one choses a four velocity vector field instead of a null one. In this case, the corresponding same-type finite relations are
\beq 
{\tilde g} = g + H \, \uu + G \, (\sun) + F \, \nn .
\eeq
The study of the signature of $ \tilde g  $ yields
$$ \begin{array}{||l|ll|}
\hline
& G^2 > (H-1)(1+F), & L,\cr
\hbox{sign($H-1$) = sign($F+1$) }=1 & G^2 = (H-1)(1+F), & D, \cr
& G^2 < (H-1)(1+F), & E. \cr
\hline
& G^2 > (H-1)(1+F), & L, \cr
\hbox{sign ($ H - 1 $) = sign ($ F + 1 $) }= -1 & G^2 = (H-1)(1+F), & D, \cr
& G^2 < (H-1)(1+F), & A.\cr
\hline
\hbox{sign ($ H - 1 $) $ \neq $ sign ($ F + 1 $)}  & L. &  \cr
\hline
\hbox{$ H = 1 $ or $ F = -1 $} & L. \hbox{ For $ G = 0 $,} & D. \cr
\hline
\hbox{$ H = 1 $, $ F = - 1 $, $ G = 0 $} & \hbox{double D.} & \cr
\hline
\hline
\end{array} $$
Now the degenerated cases are analogous to the ones presented in the two previous cases. In this set one finds some situations of interest, as the one having a double four-velocity field deformation, which should be worked with more detail, see also \cite{coll}.
\section{Conclusions}
We have devoted this work to develop a particular application of metric 
symmetries aiming at finding new solutions of Einstein's equations through 
a major control on geometrical elements of the final solutions. For instance, 
in the more considered case of \ks groups, the geometrical object is a null 
vector filed which has a clearly physical interpretation, as well. This method 
may open a new path to finding useful solutions to Einstein's equations. This 
has been shown along the text with several examples, including the resolution 
of \ks groups for Kerr-Newman \sts for their principal null directions, 
which constitute the basis for current studies on the structure of blak holes, 
or spinning particles. Moreover, this method may prove to be a complement 
of the usual (G)KS transformations, which have been considered always 
from a purely finite viewpoint only. The consequences of using \ks 
groups are still to be explored in other \sts as the ones
considered here as remarked elsewhere.

For the other cases, i.e., other new metric symmetries, the interpretations 
are still to be worked out to achieve a similar status to that of \ks symmetries. 
We have just set the basic pieces towards that goal, namely, a study regarding the signature change that may happen in the corresponding metrics. A worth control will only be gained if, at the same time, an extensive study of these new metric symmetries is carried out regarding the structure of their  Lie algebras or, at least, by finding out some relevant examples.

Finally, let us recall that other possible applications, mentioned in the introduction, i.e. that of a implementation of general metric symmetries to study energy-matter properties of \st following the line of thought of \cite{zafiris,hall3,tsampa} and the approximation to the rigidity problem, \cite{bona}-\cite{ere96},
will doubtless fill in most of the gap towards the physical comprehension of metric symmetries. Therefore, besides the mentioned general studies, these are as well  desirable areas to be further considered.  All this might lead to a feed back reaction between the algebraical and physical issues of the fabric of metric symmetries.

\section*{Acknowledgements}

I am indebted with Emili Elizalde, who has help me in the way of relating \ks symmetries with physical applications through enlightening comments and 
Bartolom\'e Coll who proposed me to consider other candidates of metric 
motions a part from the \ks ones. Alexander Ya. Burinskii is acknowledged for stimulating discussions and providing me his work prior to publication. 
S. R. H. wishes to thank the direcci\'o General de Recerca, Generalitat 
de Catalunya for financial support.

\end{document}